\documentclass{cs19proc}

\usepackage{natbib}
\usepackage{float}

\editors{G.~A. Feiden}
\publisher{Zenodo}
\conference{The 19th Cambridge Workshop on Cool Stars, Stellar Systems, and the Sun}
\conferencedate{2016}

\title{Identifying the ejected population from disintegrating multiple systems}
\author{A. K. P. Yip,$^{1,2}$ 
        D. J. Pinfield,$^{3}$ 
				R. Kurtev,$^{2,1}$ 
				M. Gromadzki,$^{1,2}$
				and F. Marocco$^{3}$}

\affiliation{$^{1}$ Millennium Institute of Astrophysics, Av. Vicua Mackenna 4860, 782-0436, Macul, Santiago, Chile \\ 
			 $^{2}$ Instituto de F\'isica y Astronom\'ia, Universidad de Valpara\'iso, Av. Gran Breta\~{n}a 1111, Playa Ancha, Casilla 5030, Valpara\'iso, Chile \\
			 $^{3}$ Centre for Astrophysics Research, School of Physics, Astronomy and Mathmatics, University of Hertfordshire, College Lane, Hatfield AL10 9AB, UK.}

\shorttitle{Disintegrating benchmark systems}
\shortauthors{A. K. P. Yip et al.}

\abs{Kinematic studies of the Hipparcos catalogue have revealed associations that are best explained as disintegrating multiple systems, presumably resulting from a dynamical encounter between single/multiple systems in the field \citep{Li2009}. In this work we explore the possibility that known ultra-cool dwarfs may be components of disintegrating multiple systems, and consider the implications for the properties of these objects. We will present here the methods/techniques that can be used to search for and identify disintegrating benchmark systems in three database/catalogues: Dwarf Archive, the Hipparcos Main Catalogue, and the Gliese-Jahrei{\ss} Catalogue. Placing distance constraints on objects with parallax or colour-magnitude information from spectrophotometry allowed us to identify common distance associations. Proper motion measurements allowed us to separate common proper motion multiples from our sample of disintegrating candidates. Moreover, proper motion and positional information allowed us to select candidate systems based on relative component positions that were tracked back and projected forward through time. Using this method we identified one candidate disintegrating quadruple association, and two candidate disintegrating binaries, all of them containing one ultra-cool dwarf.}

\begin{document}

\maketitle

\section{Introduction}
A very high fraction of stars are in binary or multiple systems. The components of these binary systems interact gravitationally with each other and with other stellar objects. This could lead to the bonds between the binary components gradually becoming weaker (as they lose binding energy), eventually causing the system to become unbound. Normally such interactions are relatively weak; binaries can interact weakly with other stars/multiples, particularly if they have a large cross-section for interaction, i.e. if they are wide binaries. Nevertheless, there are some more severe interactions between binary-star or binary-binary systems and they can lead to a more violent break-up. In theory, this could be the cause of some of the disintegrating multiple systems seen in the Hipparcos Main Catalog \citep[hereafter HMC,][]{Li2009}. Disintegrating systems are thought to be common according to \citet{Li2009}. However, since the time for the disintegration to take place is rather short \citep[$\sim 1$ Myr,][]{Szebehely1972} they can be extremely difficult to identify. Furthermore, such interactions as well as breaking up the binary components, might also release low-mass objects previously bound to one of the components. Such low-mass objects might include low-mass stars, brown dwarfs (hereafter BDs), or exoplanets. 

The type of system we are particular interested in are the ones that consist of at least one BD or ultra-cool dwarf (hereafter UCD). BDs are known as intermediate objects between stars and planets, and hold the key to bridge the gap between the two. Although BDs are faint and difficult to detect and characterize, they are far from being rare. The number ratio of BDs to stars, possibly dependent on their environment, according to various surveys vary between 1/6 and 1/3 \citep[e.g.][]{Kirkpatrick2012,Scholz2012,Luhman2007}. BDs are found either as field objects or as cluster/moving group members. They have been observed as isolated objects, as well as in binary systems (i.e. BD-BD; star-BD) and in higher order multiple systems.

The identification of disintegrating benchmark systems \citep[i.e. multiple systems consisting of UCDs and stars, e.g.][]{Pinfield2012,Day-Jones2011} would allow further testing of formation theories, e.g. turbulent fragmentation of molecular clouds \citep{HennebelleChabrier2009,HennebelleChabrier2008,PadoanNordlund2004}, disk fragmentation \citep{StamatellosWhitworth2009,Attwood2009}, and binary evolution models \citep[e.g.][]{Veras2014,Veras2011,VerasTout2012}. Furthermore, the glare of the primary star can conceal orbital populations such as giant planets, BDs and very low-mass stars, and make studies of these dimmer objects extremely challenging. Hence being able to identify systems where these low-mass objects have been revealed to us by being ejected dynamically would allow us to investigate them further. In addition, benchmark systems are fundamental to provide atmospheric parameters for UCDs \citep[e.g.][]{Pinfield2006}.

There has been some limited analysis of disintegrating multiple systems in the field using Hipparcos data \citep{Li2009}. Other related studies in the disintegrating multiple systems domain are the numerical experiments on binary-binary encounters \citep{Mikkola1983}, the dynamical evolution of star clusters \citep[e.g.][]{Aarseth2004}, focusing in particular on the slingshot mechanism as one of the formation scenario for triple systems \citep[due to binary-binary encounters;][]{Saslaw1974}, in addition to simulations on the effect that negative total energy has on triple systems \citep[e.g.][]{Anosova1990}. More recently, disintegrating multiple systems have been the focus of the post-main-sequence evolution theory of wide binaries \citep{Veras2014,Veras2011,VerasTout2012}, they have been proposed as a formation model for BD binaries \citep{ReipurthMikkola2015}, and as a possible explanation for the observed bimodal initial mass function in the Orion Nebula Cloud \citep{Drass2016}. However no actual disintegrating candidate has been identified so far, especially none that consists of at least one UCD, for us to test the models and simulations mentioned previously.

In this proceeding we present a method to search for such disintegrating systems, and our preliminary results based on the use of the HMC, the Gliese-Jahrei{\ss} Catalogue (hereafter GJC) and Dwarf Archive (hereafter DA).

\section{Method}
To search for this type of candidates associations, the first step is to cross-match various catalogues to identify objects that are close in the sky and with common distance. The three primary catalogues we used were HMC \citep{vanLeeuwen2007}, GJC \citep{Stauffer2010} and DA. DA is essential in this work as it is a catalogue of UCDs, that would allow us to identify if any of this low-mass objects are being ejected. HMC and GJC were chosen because they contain bright nearby stars that would constitute ideal benchmark systems. They also provide a good range of both spectral types and distance. 

The first step was to apply physical separation criteria. Although the furthest separations for known wide binary systems are $\sim$200 kAU \citep[e.g.][]{Caballero2010,Caballero2006}, the maximum separations observed for a main-sequence star-BD binary are $\sim$5000 AU \citep{Pinfield2006,Gizis2001}. Given we do not want to exclude possible white dwarf-BD binaries and that the separations between them could approximately be four times the common separations of main-sequence star-BD binaries \citep[i.e. $\sim$20 kAU, see e.g.][]{Day-Jones2011,Zhang2010,Faherty2010,Burgasser2005}. Considering the type of systems that we are searching for are extremely wide (because they are disintegrating) we apply a more conservative physical separation constraint of 50 kAU, to balance between reducing the number of contaminants and selecting all possible candidates. 

Then it is essential for objects in these candidate associations to be at common distance (in order to reject chance alignments). Though, many of the objects in DA do not have a measured parallax. To resolve this problem, we estimated the spectrophotometric distance for objects in the catalogue that have relatively reliable spectroscopy using the absolute magnitude-spectral type relation from \citet{DupuyLiu2012}. The quoted root-mean-square of the relation was adopted as the uncertainty. Additionally, in order to select only good quality measurements, we have imposed a limit on the uncertainty of the HMC parallaxes of 10\% or better. As a result, the catalogue was downsized by a factor of $\sim$six with only 20871 objects left.

Furthermore, it is also crucial to identify the potential causes of the disruption of the system, the most obvious of which could be the dynamical interaction between the system and a nearby object/system. Hence we cross-matched our initial candidate associations with HMC and GJC, out to a 1 degree radius from the candidate associations. This choice is based on a simple consideration. First of all, for us to be able to identify a disintegrating association, the disintegration would have to have taken place no longer than a few thousands years ago (otherwise the components would have moved too far apart). An object that would have interacted with the candidate association at the time of the disintegration, even with the highest proper motion (hereafter PM) of the systems considered, would not have had the time to travel further than 1 degree.

The next step is to separate gravitationally bound systems from unbound disintegrating associations. Widely separated bound multiple systems should have common PM (hereafter CPM), while disintegrating systems should be otherwise non-CPM. To identify CPM systems we have selected only groups where the differences between the PM of all components are within the combined error (Gaussian error). To identify disintegrating candidates, we selected only groups where the PM of at least one component of the multiple system diverges by more than three times the combined error. We note that only 35\% of the DA objects have PM measurements, and so the CPM analysis is only possible for those DA objects. Moreover, GJC does not provide errors on the PM, so we assumed an uncertainty of 2.5 mas/yr following \citet{Stauffer2010}. 

Then, as a final assessment for disintegrating multiple candidates, we determined if associated objects are moving away from each other, as we expect for disintegrating associations. The separation between associated objects was calculated as a function of time (t) using equation \ref{sept}. \\

\noindent $Sep(t) =  \sqrt{[(\alpha_1 + \mu_{\alpha_1} \times t) - (\alpha_2 + \mu_{\alpha_2} \times t)]^2 \times cos⁡(\delta_1)^2}$
\begin{equation}
 + \sqrt{[(\delta_1 + \mu_{\delta_1} \times t) - (\delta_2 + \mu_{\delta_2} \times t)]^2}
\label{sept}  
\end{equation}

Where $\alpha_1$, $\alpha_2$ and $\delta_1$, $\delta_2$ is the right ascension and the declination of the components in the association respectively, $\mu_{\alpha_1}$, $\mu_{\alpha_2}$ and $\mu_{\delta_1}$, $\mu_{\delta_2}$ is the PM in the right ascension direction and the PM in the declination direction of the components in the association respectively.

Note that in equation 1 the time can be negative and positive, reflecting the past and future separation of the association. This equation does not account for any gravitational attraction, as it assumes straight line motion for associated objects. This is an approximation, though should be reasonable for most of our candidates, because at wide separation the effects of gravity should be small. However, we expect this approximation to lead to some inaccuracies in the past/future forecasts of the time of closest approach. Therefore we expect some level of scatter in the times of closest approach for the components of higher order systems. This is because in higher order systems the gravitational interaction between the components is more complex and the uncertainties add up causing larger scatter. We have used the Monte Carlo method to calculate the uncertainty on Sep(t). 

\begin{figure*}[h]
\makebox[\linewidth]{%
\includegraphics[width=0.36\linewidth, angle=90]{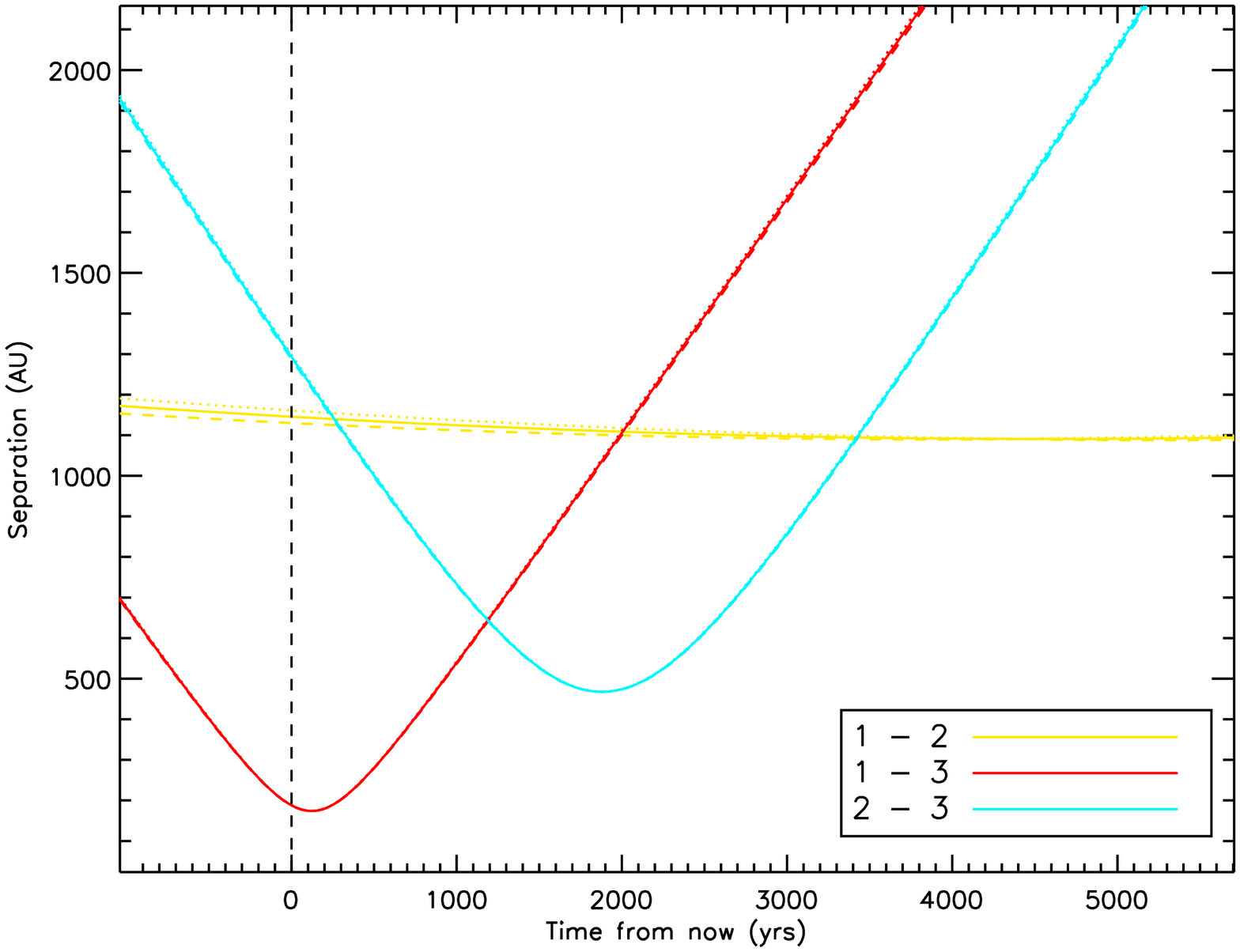}%
\hfill    
\includegraphics[width=0.36\linewidth, angle=90]{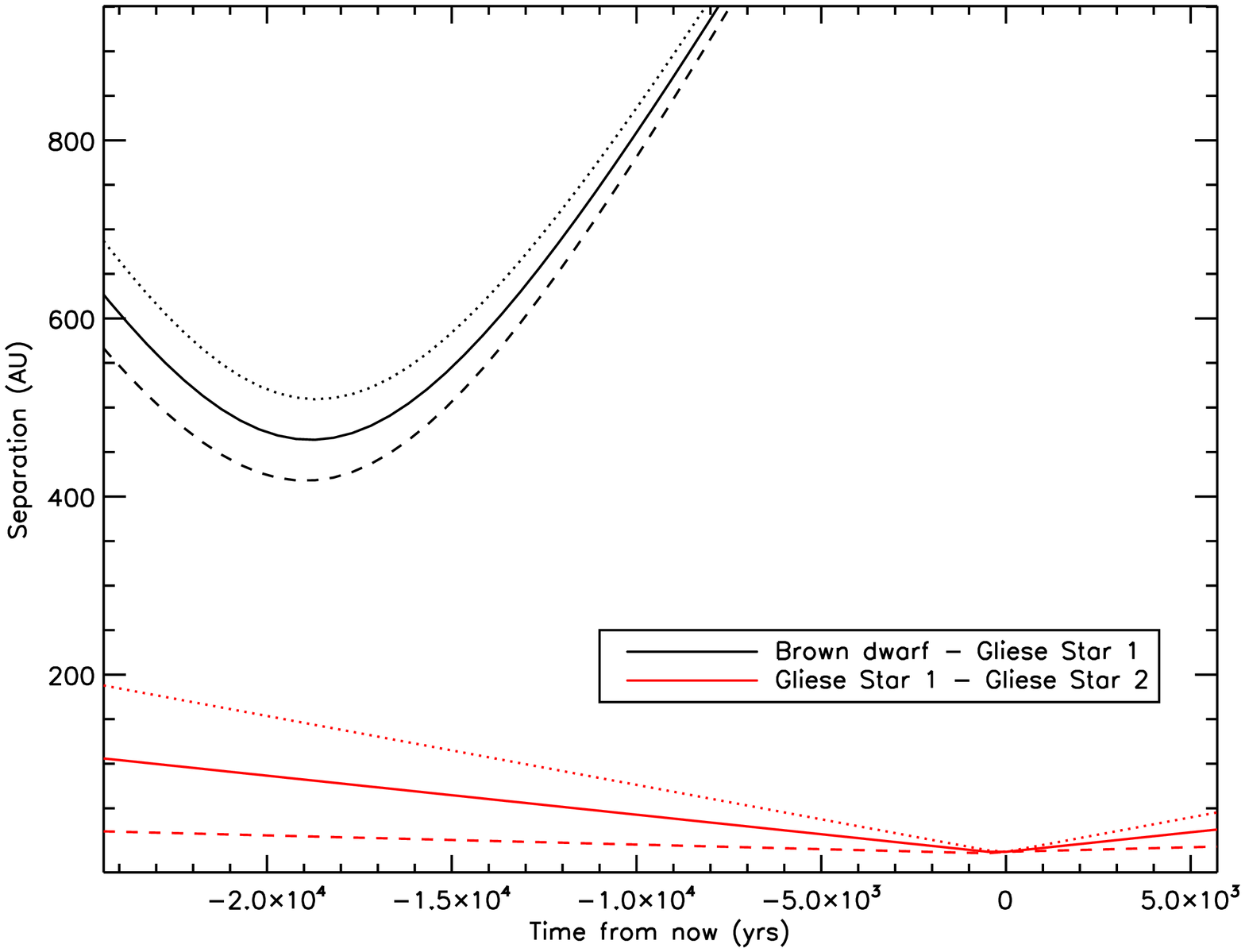}%
}
	\caption{The parabolas show the on-sky separation between the components of the associations. In the left panel, the separation between the objects in a non-CPM association as a function of time. The yellow line is the separation between object number 1 and object number 2, the red line is the separation between object number 1 and object number 3 while the blue line is the separation between object number 2 and object number 3. The objects reach minimum separation in the future so the components are moving towards each other, and the systems therefore is not disintegrating. In the right panel, the separation between the components of the association as a function of time. The black line represents the separation between the BD and the host star (``Gliese star 1'') and the red line represents the separation between ``Gliese star 1'' and the disrupting third body (``Gliese Star 2''). The association disintegrates after ``Gliese star 2'' passes between the other two objects, i.e. the minimum separation between the components in the association is in the past. For each parabola the one sigma error range is delimited by a dotted line (upper limit) and a dashed line (lower limit).}
	\label{parabola}
\end{figure*}

We then selected as disintegrating candidates only associations where the epoch of closest approach between the objects was in the past, i.e. the objects were currently dispersing. For the intermediate cases, where some objects have the closest approach in the past and some in the future, we interpreted by eye the plot and decided case by case if the system could be disintegrating or not. We show in Figure \ref{parabola} an example of a system that was classified as non-disintegrating (left panel) and a system that was classified as possibly disintegrating (right panel). As can be seen the components of the association in the left panel reach their closest approach in the future, and therefore this association is not disintegrating.  On the other hand, the association in the right panel is disintegrating after the object we refer to as ``Gliese star 2'' passes between the object we refer to as ``Gliese star 1'' and the BD. ``Gliese star 2'' continues on its path unaffected. The ejection velocity is high, as can be seen from the steep parabola, and the current separation between the BD and ``Gliese star 1'' is 1500 AU, i.e. more than 1000 AU larger than the minimum. The uncertainty on the PM of both objects is small and therefore the shape and the minimum of the parabola are very reliable.

Overall we have identified one disintegrating quadruple candidate containing a UCD component (although this is not the component with a divergent PM), as well as two double systems, each containing a UCD. However for the two double associations we could not identify a third object nearby that can cause the ejection, at least not from HMC and GJC. The first one consists of a T dwarf and a white dwarf, with the T dwarf having a PM approximately twice the magnitude of the white dwarf's ($\sim$15 sigma difference), and in essentially the same direction. The second one has been previously identified as a bound multiple system, though the PMs of its components differ by $\sim$8 sigma. In both cases the relative motions show that the components were closer together in the past, with the T-white dwarf system predicted to be within $\sim$1400 AU about 1300 years ago. Our analysis method has thus been successful in identifying possible disintegrating multiple systems that are inconsistent with gravitationally bound wide multiples. 

\section{Conclusions and future work}
A method has been defined to identify candidate disintegrating multiple systems using (i) distance constraints from trigonometric parallax measurements and spectrophotometric techniques, (ii) relative proper motions and their associated measurement uncertainties, and (iii) relative component positions that are tracked back and projected forward through time. This method has been applied to three catalogue/database compilations, namely DA (for UCDs), HMC, and GJC. The robustness of the method is proven by the fact that we have been able to identify many previously confirmed binaries, multiples, benchmark systems and cluster objects. 

Confirmation and further study of our candidate disintegrating multiple associations should be able to place new constraints on the typically unseen constituents of multiple systems, by revealing the full range of components during disintegration. The observed frequency of disintegrators will also inform us about the rate at which such systems interact in the Galactic disk. And through further work we should be able to extend the range over which we can identify disintegrating components to lower masses.

The candidate disintegrating systems identified in this work should be examined further using \emph{Gaia} \citep{CacciariPancinoBellazzini2015} data and ground-based follow-up observations. More accurate PM and parallax measurements of the components would assess our own selection criteria at higher accuracy. It would be particularly interesting to examine early \emph{Gaia} PMs to confirm if our disintegrating quadruple association really does have a member with divergent proper motion. Radial velocity measurements (from \emph{Gaia} or ground-based telescopes) would also give 3-D velocities for the components of our candidate associations, which would allow dynamical studies to determine if the systems are gravitationally unbound. If the T dwarf-white dwarf disintegrating double system is confirmed by \emph{Gaia}, then a detailed study could constrain the white dwarf cooling age, and search for any unusual features of the T dwarf spectrum that may have arisen during a period when the components were in closer proximity. 

Beyond our current analysis of the DA, HMC and GJC, we can also expand the search to additional surveys. \emph{Gaia} will soon provide extremely accurate PMs and parallaxes for stars and UCDs down to V=20. And a much larger sample of UCDs could be included by selecting photometric candidates from UKIDSS, UHS, VHS, WISE and SDSS.

The contribution in searching for possible disintegrating multiple systems in this work is fundamental in providing further constrains to the formation models. Understanding the dominant formation mechanism of ultra-cool objects have been for a long time one of the main research interests among the low-mass community. Various formation models have been suggested \citep[e.g.][]{Stamatellos2012}, but to improve these models, observational constrains are fundamental, especially on the IMF and the binary fraction. This work might provide an answer to the origin of some of the field object and even the initial requirements for capture.

\section*{Acknowledgments}
{This research has benefitted from the M, L, T, and Y dwarf compendium housed at DwarfArchives.org. AY would like to thank the Cool Stars 19 organising committee and Proyecto CONICYT-REDES140024, ``SOCHIAS grant through ALMA/Conicyt Project \#31150039'' and Ministry of Economy, Development, and Tourism's Millennium Science Initiative through grant IC120009, awarded to The Millennium Institute of Astro- physics, MAS for financial support.}

\bibliographystyle{cs19proc}
\bibliography{ayip_proceeding}

\end{document}